\documentclass[12pt]{article}

\usepackage{amsmath}
\usepackage{amssymb}

\newcommand{\la}{\lambda}
\newcommand{\K}{\mathrm{K}}

\newcommand{\om}{\omega}

\begin{document}

\title{Dissipationless shock waves in repulsive Bose-Einstein condensates}

\author{A.M. Kamchatnov$^1$, A. Gammal$^{2}$, and
R.A. Kraenkel$^3$\\
$^1$Institute of Spectroscopy, Russian Academy of Sciences,\\ Troitsk 142190,
Moscow Region, Russia\\
$^2$Instituto de F\'{\i}sica, Universidade de S\~{a}o Paulo,\\
05315-970, C.P.66318 S\~{a}o Paulo, Brazil\\
$^3$Instituto de F\'{\i}sica Te\'{o}rica, Universidade Estadual Paulista-UNESP\\
Rua Pamplona 145, 01405-900 S\~{a}o Paulo, Brazil\\}

\maketitle

\begin{abstract}
We consider formation of dissipationless shock waves in Bose-Einstein
condensates with repulsive interaction between atoms. It is shown
that big enough initial inhomogeneity of density leads to wave breaking
phenomenon followed by generation of a train of dark solitons.
Analytical theory is confirmed by numerical simulations.
\end{abstract}

Experiments on free expansion of Bose-Einstein condensate (BEC) have
shown \cite{ernst} that evolution of large and smooth
distributions of BEC is described very well by hydrodynamic
approximation \cite{cd} where dispersion and dissipation effects are
neglected. At the same time, it is well known from classical
compressible gas dynamics (see, e.g. \cite{LL}) that typical initial
distributions of density and velocity can lead to wave breaking phenomenon
when formal solution of hydrodynamical equations becomes multiple-valued. It
means that near the wave breaking point one cannot neglect dispersion
and/or dissipation effects which prevent formation of a multiple-valued
region of a solution. If the dissipation effects dominate the
dispersion ones, then the multiple-valued region is replaced by the
classical shock, i.e. narrow layer with strong dissipation within, which
separates smooth regions with different values of density, fluid
velocity and other physical parameters. This situation was studied in
classical gas dynamics and found many practical applications. If,
however, the dispersion effects dominate dissipation ones, then the
region of strong oscillations is generated in vicinity of wave
breaking point \cite{GP,kamch}. Observation of dark
solitons in BEC \cite{dark1,dark2,dark3} shows that
the main role in dynamics of
BEC is played by dispersion and nonlinear effects taken into
account by standard Gross-Pitaevskii (GP) equation \cite{PS}, and
dissipation effects are relatively small and can be considered as
perturbation. Hence, there are initial distributions of
BEC which can lead to formation of dissipationless shock waves.
Here we shall consider such  possibility.

The starting point of our consideration is the fact that the
sound velocity  in BEC is proportional to the square root from
its density (see, e.g. \cite{PS} and references therein).
Thus, if we create inhomogeneous BEC with high density hump (with
density $\sim\rho_1$) in the center of lower density distribution
(with density $\sim\rho_0$), and after that release this central
part of BEC, then the high density hump will tend to expand with
velocity $\sim\sqrt{\rho_1}$ greater than the sound velocity
$\sim\sqrt{\rho_0}$ of propagation of disturbance in lower density
BEC. As a result, wave breaking and formation of dissipationless
shock wave can occur in this case. Note that initial distribution
of this type was realized in the recent experiment \cite{cornell}
where generation of oscillations was also observed. In this
experiment the lower density disk-shaped BEC was confined by magnetic trap
and density distribution had standard Thomas-Fermi (TF) parabolic
form. Additional potential was applied to BEC in the central part of
above TF profile which lead to narrow parabolic hump in the
density distribution. After the central potential was switched off,
the hump started to expand against wide lower density TF profile
leading to generation of oscillations in the transition region between
high and low density condensates. The theory of dissipationless
shock waves in media described by one-dimensional (1D) nonlinear
Schr\"{o}dinger (NLS) equation was developed in \cite{kamch01,kku02}.
Since the GP equation in some cases can be reduced to the 1D NLS
equation, this theory can be applied to description of dissipationless
shock waves in BEC. Here we suggest such
description and confirm it by numerical simulations.

We consider BEC confined in a ``pancake'' trap with the axial
frequency $\omega_z$ much greater than the transverse one
$\omega_x=\omega_y=\omega_\bot$.
Let the density of atoms in the central part of BEC be of order
of magnitude $n_0$ and satisfy the condition $n_0a_sa_z^2\ll1$,
where $a_s>0$ is the $s$-wave scattering length and $a_z=
(\hbar/ma_z)^{1/2}$ is the amplitude of quantum oscillations
in the axial trap. Then the condensate
wavefunction $\psi$ can be factorized as $\psi=\phi(z)\Psi(x,y)$,
where $\phi(z)=\pi^{-1/4}a_z^{-1/2}\exp(-z^2/2a_z^2)$ is the
ground state wavefunction of axial motion, and $\Psi(x,y,t)$
satisfies the reduced 2D GP equation
\begin{equation}\label{GP}
i\hbar\Psi_t=-\frac{\hbar^2}{2m}(\Psi_{xx}+\Psi_{yy})+V(x,y)\Psi
+g_{2D}|\Psi|^2\Psi,
\end{equation}
where $V(x,y)$ is the potential of a transverse trap,
$g_{2D}=2\sqrt{2\pi}\hbar^2a_s/(ma_z)$ is the effective nonlinear
interaction constant, and $\Psi$ is normalized to the number of
atoms, $\int|\Psi|^2dxdy=N$. The initial distribution of density is
determined by the potential $V(x,y)$ and consists of wide
background with a hump in its center. We assume that the background
width is much greater than the hump's width, so that at the initial
stage of evolution we can consider an expansion of the central part
against the constant background. In a similar way, at the initial
stage of evolution, when the radius of the central part does not
change considerably, we can neglect the curvature of axially
symmetrical distribution and consider its 1D cross section. It
means that we can neglect the dependence of $\Psi$ on $y$
coordinate and consider $\Psi$ as a function of $x$ and $t$
only. As a result, we arrive at 1D NLS equation with
inhomogeneous initial distribution of density. To simplify the
notation, we introduce dimensionless variables $t'=\om_z t/2$,
$x'=x/a_z,$ $u=(2\sqrt{2\pi}a_sa_z/n_0)^{1/2}\Psi$.
Then the initial stage of evolution of the wavefunction profile in
the $x$ axis cross section is governed by the NLS equation,
\begin{equation}\label{NLS}
iu_{t}+u_{xx}-2|u|^2u=0,
\end{equation}
where the primes in $t'$ and $x'$ are omitted for convenience of
the notation.

Evolution of smooth pulses before the wave breaking point can be
described in the hydrodynamic approximation which can be achieved
by substitution
\begin{equation}\label{madelung}
u(x,t)=\sqrt{\rho(x,t)}\exp\left(i\int^xv(x',t)dx'\right)
\end{equation}
and separation of the real and imaginary parts. As a result we obtain
the system
\begin{equation}\label{euler}
\tfrac12\rho_t+(\rho v)_x=0,\quad
\tfrac12v_t+vv_x+\rho_x=0,
\end{equation}
where we have neglected the so-called ``quantum pressure'' term with
higher space derivatives what is correct until the density
distribution has smooth enough profile.
To get simple qualitative
picture of the wave breaking of BEC density distribution, let us
consider idealized case with a box-like hump in the initial
distribution,
\begin{equation}\label{hump}
\rho(x,0)=\Bigg\{
\begin{array}{l}
\rho_0,\quad |x|>a,\\
\rho_1,\quad |x|\leq a.
\end{array}
\end{equation}
Although this distribution has a parameter with dimension of
length---width of the hump---it does not play any role for
$t\leq a/(2\sqrt{\rho_1})$, i.e. until two waves propagating inward
the hump with sound velocity $c_1=2\sqrt{\rho_1}$ meet at $x=0$.
Hence, for this initial period of evolution the solution for
$x>0$ can only depend on the self-similar variable $\xi=(x-a)/t$,
centered at $x=a$, and for $x<0$ on $\xi=(x+a)/t$ centered at
$x=-a$. Since the picture is symmetrical, it is enough to
consider only a half of the solution corresponding to $x>0$.
It is easy to find that the density is given by the formulae
\begin{equation}\label{self}
\rho(x,t)=\left\{
\begin{array}{ll}
\rho_1 & \quad\mathrm{for}\quad 0<x<a-2\sqrt{\rho_1}\,t,\\
\left[2\sqrt{\rho_1}-(x-a)/2t\right]^{2}/9
& \quad\mathrm{for}\quad
a-2\sqrt{\rho_1}t<x<a+2(\sqrt{\rho_1}-\sqrt{\rho_0})t,\\
\left[2\sqrt{\rho_0}+(x-a)/2t\right]^2/9
& \quad\mathrm{for}\quad a+2\sqrt{\rho_0}\,t<x<
2+2(\sqrt{\rho_1}-\sqrt{\rho_0})t,\\
\rho_0 & \quad\mathrm{for}\quad x>a+2\sqrt{\rho_0}\,t,
\end{array}
\right.
\end{equation}
and similar formulae can be written for the velocity field
$v(x,t)$. This solution describes the wave breaking phenomenon
which takes place if $\rho_1>4\rho_0$. The density profile shown in
Fig.~1 clearly illustrates the origin of the multiple-valued region
which should be replaced by the oscillatory shock wave when
the dispersion effects are taken into account.

For more realistic initial pulses the density profile is
a smooth function without cusp points. In vicinity of the wave
breaking point the solution can be approximated by a cubic function
for one Riemann invariant $\lambda_+=v/2+\sqrt{\rho}$ of the
system (\ref{euler}) and by constant value for
another one  $\lambda_-=v/2-\sqrt{\rho}$ (see \cite{kamch01,kku02}).
After Galileo and scaling transformations the hydrodynamic
solution can be written in the form
\begin{equation}\label{hydro}
x-(3\lambda_++\lambda_-)t=-\lambda_+^3,\quad \lambda_-=
\mathrm{const},
\end{equation}
and again for $t>0$ it has a multiple-valued region
of $\lambda_+$. It means that dispersion effects have to
be taken into account which lead to formation of dissipationless
shock wave after wave breaking point.

In framework of
Whitham theory of modulations \cite{whitham,kamch}
one can obtain an approximate solution of the NLS
equation (\ref{NLS}) in analytic form where the dissipationless
shock wave is presented as a modulated periodic nonlinear wave
solution of the NLS equation.
The density is expressed in terms of Jacobi elliptic function
\begin{equation}\label{per}
\rho(x,t)=|u(x,t)|^2=\tfrac14(\lambda_1-\lambda_2-\lambda_3+
\lambda_4)^2+(\lambda_1-\lambda_2)(\lambda_3-\lambda_4)
\mathrm{sn}^2[\sqrt{(\lambda_1-\lambda_3)(\lambda_2-\lambda_4)}
\xi,m],
\end{equation}
where
\begin{eqnarray}
\xi=x-(\lambda_1+\lambda_2+\lambda_3+\lambda_4)t,\qquad
m=\frac{(\lambda_1-\lambda_2)(\lambda_3-\lambda_4)}
{(\lambda_1-\lambda_3)(\lambda_2-\lambda_4)},
\end{eqnarray}
and parameters $\lambda_i$, $i=1,2,3,4,$ change slowly along the
dissipationless shock. Their dependence on $x$ and $t$ is
determined implicitly by the solution
\begin{equation}\label{hod}
x-v_i(\lambda)t=w_i(\lambda),\quad i=1,2,3;\qquad
\la_4=\overline{\lambda}=\mathrm{const}
\end{equation}
of Whitham equations,
where Whitham velocities $v_i$ and $w_i$ are given by quite complicated
expressions in terms of elliptic integrals (see \cite{kamch01,kku02}):
\begin{equation}
\label{solut}
w_i=-\tfrac{8}{35}w_i^{(3)}+\tfrac45\overline{\la}w_i^{(2)}-
\tfrac1{35}\overline{\la}^2v_i(\la)+\tfrac1{35}\overline{\la}^3,
\quad i=1,2,3,
\end{equation}
\begin{equation}
w_i^{(k)}=W^{(k)}+(v_i-s_1)\partial_iW^{(k)},
\end{equation}
\begin{equation}
W^{(1)}=V=s_1,\quad W^{(2)}=\tfrac38s_1^2-\tfrac12s_2,\quad
W^{(3)}=\tfrac5{16}s_1^3-\tfrac34s_1s_2+\tfrac12s_3,
\end{equation}
\begin{equation}
\label{v}
v_i(\la)=\left(1-\frac{L}{\partial_iL}\partial_i\right)V,\quad
\partial_i\equiv\partial/\partial \la_i,\quad i=1,2,3,4,
\end{equation}
where
\begin{equation}
\label{L}
L=\frac{{\rm K}(m)}{\sqrt{(\la_1-\la_3)(\la_2-\la_4)}}
\end{equation}
is a wavelength,
${\K}(m)$ is the complete elliptic integral of the first
kind, and $s_1,$ $s_2,$ $s_3$ are determined by the
expressions
\begin{equation}
s_1=\sum_{i}\la_i, \quad
s_2=\sum_{i<j}\la_i\la_j,\quad
s_3=\sum_{i<j<k}\la_i\la_j\la_k.
\end{equation}
Equations (\ref{solut}) can be solved with respect to $\la_i,$
$i=1,2,3,$ giving them as functions of $x$ and $t$.
Subsequent substitution of these functions $\la_i(x,t)$,
$i=1,2,3,$ into (\ref{per}) yields the modulated periodic
wave which represents the dissipationless shock wave.
The resulting profile of density in dissipationless shock wave
is shown in Fig.~2. At one its edge it consists of
the train of dark solitons, and at another edge describes small
amplitude oscillations propagating with local sound velocity
into unperturbed region described by smooth solution of
hydrodynamical equations.
The modulated periodic wave replaces the multiple-valued region
shown by dashed line which was obtained in hydrodynamic
approximation after the wave breaking point. This multiple-valued
region is analogous to that in Fig.~1 with account of change
of variables due to Galileo and scaling transformations.

To check the described above picture of formation of dissipationless
shock wave and to extend it to real 2D situation, we have solved
numerically the 2D generalization of Eq.~(\ref{NLS}),
\begin{equation}\label{2D-NLS}
iu_t+u_{xx}+u_{yy}-2|u|^2u=0,
\end{equation}
with the initial condition
\begin{equation}\label{parabola}
\rho(r,0)=\Bigg\{
\begin{array}{ll}
\rho_0+(\rho_{1}-\rho_0)(1-r^2/a^2), & \quad |r|\leq a,\\
\rho_0, & \quad |r|>a,
\end{array}
\end{equation}
(where $r=(x^2+y^2)^{1/2}$)
similar to one studied experimentally \cite{cornell}.
Plot of two-dimensional initial density distribution is shown
in Fig.~3 and density distribution after time evolution
$t=2$ is shown in Fig.~4.
As we see, the parabolic hump expands with formation of
dissipationless shock wave in the transition layer between
high density region to low density one. To see more
clearly the evolution of the hump,  its
cross section profiles are shown in Fig.~5 at different values of
time $t$. Slowly propagating
dark solitons are clearly seen as well as small amplitude
sound waves propagating into undisturbed low density region.
Dissipationless shock wave generated at the right side of
the profile coincides qualitatively with results of
analytic theory  shown in Fig.~2. One may assume that
oscillations in BEC density profile observed in
experiment \cite{cornell} have the same origin.

Further evolution of the hump will ultimately lead to its
spreading over large area with small amplitude oscillations,
that is the shock wave does not persist permanently.
Rather, it is a transient phenomenon caused by different
values of characteristic velocities in high density
hump and low density background. Slowly propagating
dark solitons are generated in the transient layer
between these two regions in order to reconcile
two different values of velocities of disturbance
propagation. This mechanism of dark soliton generation
is quite general and can manifest itself in various
geometries and initial BEC distributions
\footnote{Formation of dark soliton trains in 1D geometry was considered
in recent preprint \cite{Damski}}.

In conclusion, we have studied theoretically and numerically the
process of formation of dissipationless shock waves in
the density distribution of BEC. We believe that
oscillations in BEC density profile observed in recent
experiment \cite{cornell} can be explained by this
theory.

This work was supported by FAPESP (Brazil) and CNPq (Brazil).
A.M.K. thanks also RFBR (grant 01--01--00696) for partial support.

\newpage

\centerline{\bf Figures captions}

\bigskip
\bigskip

Fig.~1. Wave breaking of a box-like initial density distribution (shown by
dashed line) in hydrodynamic approximation. The plot corresponds to
the solution (\ref{self}) with $a=10,$ $\rho_0=1,$ $\rho_1=10,$
$t=1$. Point $A$ propagates inward the box with local sound
velocity $2\sqrt{\rho_1}$, point $C$ propagates outward along
the background density with local sound velocity $2\sqrt{\rho_0}$,
and point $B$ corresponds to intersection of two simple wave
solutions with profiles $AB$ and $CB$.

\bigskip

Fig.~2.
Formation of dissipationless shock wave after wave breaking
point according to Whitham modulation theory applied to
1D NLS equation. Dashed line corresponds to multiple-valued region
arising in hydrodynamic approximation given by Eq.~(\ref{hydro})
(it is analogous to the region between the points B and C in Fig.~1),
and solid line
represents modulated periodic wave given by Eqs.~(\ref{per})
and (\ref{hod}). Both profiles are calculated for $\overline{\la}=-10$
at time $t=1$.

\bigskip

Fig.~3.
Two-dimensional initial distribution of BEC density with paraboloid
hump on constant background given by Eq.~(\ref{parabola})
with $a=10,$ $\rho_0=1,$ $\rho_1=10$.

\bigskip

Fig.~4.
Two-dimensional density distribution of BEC after time $t=2$ of
evolution from initial paraboloid density on constant background
according to numerical solution of 2D NLS equation.

\bigskip

Fig.~5.
Cross sections of density profile at different evolution time
according to numerical solution of 2D NLS equation (\ref{2D-NLS})
with initial condition (\ref{parabola}).

\end{document}